\documentclass[twocolumn,trackchanges]{aastex7}

\usepackage{booktabs}
\usepackage{multirow}
\DeclareUnicodeCharacter{2212}{-}

\begin{document}

\title{Deciphering Profile Stability in Millisecond Pulsars: Timescales, Frequency Evolution, and Implications on Emission Mechanisms}

\author[0009-0002-3211-4865]{Ankita Ghosh}
\affiliation{National Centre For Radio Astrophysics, Tata Institute of Fundamental Research, Pune 411007, India}
\email{ankita@ncra.tifr.res.in}

\author[0000-0002-6287-6900]{Bhaswati Bhattacharyya}
\affiliation{National Centre For Radio Astrophysics, Tata Institute of Fundamental Research, Pune 411007, India}
\email{bhaswati@ncra.tifr.res.in}

\author[0009-0001-9428-6235]{Rahul Sharan}
\affiliation{National Centre For Radio Astrophysics, Tata Institute of Fundamental Research, Pune 411007, India}
\email{rsharan@ncra.tifr.res.in}

\author[0000-0003-2122-4540]{Patrick Weltevrede}
\affiliation{Jodrell Bank Centre for Astrophysics, Department of Physics and Astronomy, The University of Manchester, UK}
\email{patrick.weltevrede@manchester.ac.uk}

\author[0000-0002-2892-8025]{Jayanta Roy}
\affiliation{National Centre For Radio Astrophysics, Tata Institute of Fundamental Research, Pune 411007, India}
\email{jroy@ncra.tifr.res.in}

\author[0000-0002-3764-9204]{Sangita Kumari}
\affiliation{National Centre For Radio Astrophysics, Tata Institute of Fundamental Research, Pune 411007, India}
\email{skumari@ncra.tifr.res.in}

\begin{abstract}
Pulse profile stability in millisecond pulsars (MSPs) is a key factor in achieving high-precision timing essential for detecting nanohertz gravitational waves with Pulsar Timing Arrays (PTAs). In this work, we present a systematic analysis of profile stabilization timescales in MSPs using a direct method based on pulse stacking, applied to long-term multi-epoch observations. Our study utilizes data from the upgraded GMRT (uGMRT) between 300--750 MHz for nine MSPs over 3–5 years and Parkes Ultra-Wideband, low-frequency receiver observations (Parkes UWL; covering 704--4032 MHz) for three of them. We find that stable profiles typically require averaging over 10$^5$--10$^6$ pulses. This is the first time such a quantitative approach has been applied to MSPs across a wide frequency range, providing an indirect but practical estimate of jitter noise -- a dominant noise source in PTA datasets. We observe that stabilization timescales depend on signal-to-noise ratio, pulse morphology, and surface magnetic field strength, with a moderate correlation indicating a possible role of magnetic field in emission stability. A complementary single-epoch analysis of nine bright MSPs with uGMRT Band-3 (300--500 MHz) reinforces these results and demonstrates the method’s applicability to broader MSP populations. We show that a strong correlation exists between profile‑stability slope and the jitter parameter, implying that, for faint MSPs profile‑stability analysis can act as an effective proxy for intrinsic pulse‑shape variability. Our work provides a novel and scalable framework to assess intrinsic profile variability, helping to guide integration time choices and reduce timing noise in PTA experiments. 

\end{abstract}

\section{Introduction} \label{sec:introduction}
Radio emissions from pulsars vary significantly in pulse shape and intensity. However, averaging hundreds to thousands of pulses produces a stable mean profile. The timescale required to achieve this stability differs among pulsars and depends on their characteristics. Millisecond pulsars (MSPs, \citet{1982Natur.300..728A}) exhibit highly stable rotational behavior, making them ideal for precision timing. Studies on MSPs have primarily focused on intrinsic pulse-to-pulse variations caused by random phase jitter \citep{1985ApJS...59..343C}, which can introduce timing uncertainties. The first systematic study of profile stability conducted by \citeauthor{1975ApJ...198..661H} (1975, hereafter HMT75) examined 12 normal pulsars. They found that pulsars with sub-pulse drifting stabilized faster, whereas those exhibiting nulling or mode changing required significantly more time to reach stability.\par
Building on this work, \citet{1995ApJ...452..814R} (hereafter RR95) analyzed a larger sample of 28 normal pulsars, including those from HMT75. 
Their findings aligned with HMT75, showing that the stability rate of normal pulsars remains largely constant over several years. They concluded that pulse profile formation is influenced by factors beyond purely stochastic noise.
Even though studies similar to HMT75 and RR95 has not been conducted for MSPs, pulse-to-pulse shape variation is investigated in a number of MSPs \citep{2011MNRAS.418.1258O,2012MNRAS.420..361L, 2014MNRAS.443.1463S,2016ApJ...819..155L,2021MNRAS.502..407P, 2024ApJ...964....6W}, allowing them to place constraints on the resulting jitter noise. In this paper, rather than constraining the jitter noise directly, we focus on the stability of the pulse profile as a function of integration time, and explore how this stability correlates with other pulsar parameters. The profile stability is naturally influenced by jitter noise, i.e., underlying pulse-to-pulse shape variations. Given the challenges in isolating jitter noise for MSPs with typical brightness, we instead characterize its impact indirectly by determining the profile stability timescale, beyond which integrated profiles remain stable and timing precision improves. Thus, we argued that this method is particularly beneficial for weaker MSPs.\par
Profile stability analysis provides insights into the emission physics of MSPs, including how magnetospheric processes shape pulse profiles \citep{1998ApJ...501..270K}. By studying how profiles stabilize over time, one can distinguish between intrinsic pulsar properties and external influences such as interstellar scattering \citep{2004ASPC..317..211C}. Another major aim of this work is to explore whether the profile stability rate correlates with intrinsic pulsar parameters such as characteristic age, spin period, period derivative, and magnetic field strength. Identifying such trends can offer insight into the physical processes driving pulse shape fluctuations and emission stability in millisecond pulsars.
Motivated by this, we present an analysis of profile stability rates for a sample of nine MSPs discovered by GMRT using several years of observations from the uGMRT. We also examine stability characteristics of nine other bright MSPs observed in a single epoch using the uGMRT band--3. Section \ref{sec:Observations} details the observations and methodology used for profile stability analysis. In Section \ref{sec:results}, we examine the correlation between profile stability rates and various pulsar properties. Finally, Section \ref{sec:summary and conclusions} summarizes our key findings and conclusions.

\section{Observations and profile stability analysis} \label{sec:Observations}

\begin{center}
\renewcommand{\tabcolsep}{2.5pt} 
\begin{table*}[!htb]
\begin{center}
{\footnotesize
\caption{Parameters for profile stabilization. }
\label{tab:1}
\begin{tabular}{lcccccccccc} 
\hline
\toprule
MSP & P & $\dot{P}_{-21}$ & S400$^{\dag}$ & Pulses$^{\ddagger}$  & $n$ & $S_{1}$ & $S_{2}$  &$\alpha_n$ \\  
& (ms) & &(mJy) & $\times 10^{5}$ & $\times 10^{4}$ & & & & & \\
\midrule
\multicolumn{11}{c}{GMRT discovered MSPs with several years of observations}\\
\hline

J0248$+$4230$^{*}$ & 2.601 & 16.873 & & 12.5 & 3.2 & --0.14  & ---0.90(3) & 1.3 (6)  \\
J1120$-$3618 & 5.557 & 0.875 & 3.24 & 6.0 & 0 & \multicolumn{2}{c}{--0.84(12)}  & 1.3(1) \\
J1242$-$4712$^{**}$ & 5.313 & 21.445 & 2.0 & 4.5 & 0 & \multicolumn{2}{c}{--1.36(13)}  & 1.4(2) \\
J1536$-$4948 & 3.080 & 21.189 & 4.79 & 5.7 & 0 & \multicolumn{2}{c}{--1.18(7)} & 1.9(5) \\
J1544$+$4937$^{**}$ & 2.159 & 2.796 & 5.4 & 18.7  & 13.1 & --0.23(6) & --0.83(7)  & 1.6(6) \\
J1646$-$2142 & 5.853 & 8.307 & 1.58 & 4.6   & 0 & \multicolumn{2}{c}{--0.60(3)} & 1.2(3)\\
J1828$+$0625 & 3.628 & 4.688 & 8.31 & 6.1   & 1.6 & --0.22(8) & --0.85(5)  & 1.3(3) \\
J2101$-$4802 & 9.480 & 15.811 & 10.2 & 2.3    & 0.8 & --0.10 & --1.003(17)  & 2.0(3) \\
J2144$-$5237 & 5.041 & 9.056 & 1.23 & 6.3    & 0 & \multicolumn{2}{c}{--0.48(8)} & 2.3(2) \\
 
\midrule
\multicolumn{11}{c}{Bright MSPs with a single epoch observation}\\
\hline

J0218$+$4232 & 2.323 & 77.395 & 47 &  6.3  & 3.2 & --0.95 & --0.31(1) & ...\\
J0437$-$4715 & 5.757 & 57.292 & 550 & 2.6   & 0 & \multicolumn{2}{c}{--0.75(3)} & ... \\
J1022$+$1001 & 1.645  & 43.337  & 75  &  0.9  & 0.4  & --0.03 & --0.37(7) & ...\\
J1640$+$2224 & 3.163 & 2.817 & 8 & 5.6 & 3.3 & --0.85 & --0.68(4) & ... \\
J1713$+$0747 & 4.570 & 8.529 & 6.8 & 3.8  & 0 & \multicolumn{2}{c}{--1.05(2)} & ... \\
J1908$+$2105$^{**}$ & 2.564 & 13.835 & 6.45 & 6.8 & 3.2 & --0.60  & --1.02(5) & ...\\
J1909$-$3744 & 2.947 & 14.025 & 15.14 &  3.9  & 1.6 & --2.20 & --1.04(1) & ... \\
J2124$-$3358 & 4.931 & 20.568 & 17 & 2.7  & 1.6 & --0.74 & --1.01(5) & ... \\
J2145$-$0750 & 1.605 & 29.792 & 46 & 0.8  & 0 & -2.63 &--0.8 (3) & ... \\

\hline
\end{tabular}
}
\end{center}

{\footnotesize {{\bf Notes.}}
$^{*}$Isolated MSP, $^{**}$spider binaries; rest are wider MSP binaries\\
$^{\dag}$The flux density at 400 MHz for PSRs J1242$-$4712 and J2101$-$4802 is adopted from \citet{https://doi.org/10.3847/1538-4357/acc10f}, while for all other pulsars it is obtained from the ATNF Pulsar Catalogue and extrapolated using the spectral indices reported by \citet{https://doi.org/10.1017/pasa.2022.19}.\\
$^{\ddagger}$Average number of pulses in each observation\\
For MSPs with no detected slope break, the single measured slope is reported centered across both $S_1$ and $S_2$ columns.
}
\end{table*}

\end{center}

Observations for the nine GMRT-discovered MSPs were taken over several years with the uGMRT \citep{2017CSci..113..707G} band--3 (300--500 MHz) and band--4 (550--750 MHz), while a single epoch of observations was used for the nine other bright MSPs. We folded each epoch using the pulsar's parameter file with {\sc{PRESTO}}'s \citep{2002AJ....124.1788R} ``{\sc{prepfold}}'', obtaining profiles with 60 sub-integrations and 64 phase bins per period. We also observed PSR J1242--4712, PSR J2101--4802, and J2144--5237 using Parkes UWL receiver \citep{2020PASA...37...12H}, with frequency ranging from 704--4032 MHz. The data were coherently de-dispersed, folded using the pulsar's topocentric periodicity, and divided into 30-second subintegrations with 64 phase bins per period. A baseline-subtracted total intensity was determined by removing the off-pulse phase's average intensity, and a global average profile was obtained by synchronously averaging all pulses.\par
To obtain the timescale over which the profile stabilizes, we followed the approach outlined by HMT75 and RR95. Sets of subaveraged profiles were obtained over $n$ = 2$^{m}$ adjacent pulses, in which $m$ = 0 to $l$ such that $2^{l}$ is less than one-half the total number of periods. 
Cross-correlation coefficients $X^{k}_{n}$ between the subaveraged profile and the global profile (indicating the profile stability) were obtained, where $k$ is the number of subaverages for each value of $n$ ($k$ $=$ integer$(N/ 2^{m})$, where $N$ is the total number of pulses observed). The average cross-correlation coefficient $X_{n}$, was then obtained by averaging over all the $k$-independent subaverages. To determine the average correlation for a given MSP over multiple epochs, the correlation coefficients for all subaverages ($X_{n}$) computed for each epoch were averaged across all observations. The error bars represent the standard deviation of the correlation coefficients \( X_{n} \). The behavior of ($1-X_{n}$) with $n$ gives a measure of the rate of profile stabilization (Figure \ref{fig:RR}).  The mathematical formalism for this analysis is detailed below.\par
To quantify profile stability, we analyze the evolution of the pulse shape as a function of the number of averaged pulses. Let $I(\phi, t)$ represent the intensity of the pulsar signal as a function of pulse phase $\phi$ and time $t$, with $N_\phi$ phase bins per pulse period. The global average profile is defined as:

$$
I_{\text{g}}(\phi) = \langle I(\phi, t) \rangle_{t,\nu}
$$

where the averaging is performed over all available time samples and frequency channels after baseline removal. We construct subaveraged profiles by averaging over $n$ consecutive pulses starting at time $t_0$,

$$
I_n(\phi, t_0) = \frac{1}{n} \sum_{i=1}^{n} I(\phi, t_0 + i \cdot P)
$$

where $P$ is the pulsar period. Following the approach of HMT75 and RR95, we scale each subaveraged profile to match the total intensity of the global profile,

$$
\alpha_n = \frac{\sum_{\phi=1}^{N_\phi} I_{\text{g}}(\phi)}{\sum_{\phi=1}^{N_\phi} I_n(\phi, t_0)} \quad ; \quad I_{n,\text{scaled}}(\phi, t_0) = \alpha_n \cdot I_n(\phi, t_0)
$$

We then quantify the similarity between the scaled subaverage and the global profile using the Pearson correlation coefficient as follows,

$$
X_n(t_0) = \frac{\sum_{\phi=1}^{N_\phi} [I_{\text{g}}(\phi) - \bar{I}_{\text{g}}][I_{n,\text{scaled}}(\phi, t_0) - \bar{I}_{n,\text{scaled}}]}{\sqrt{\sum_{\phi=1}^{N_\phi} [I_{\text{g}}(\phi) - \bar{I}_{\text{g}}]^2 \sum_{\phi=1}^{N_\phi} [I_{n,\text{scaled}}(\phi, t_0) - \bar{I}_{n,\text{scaled}}]^2}}
$$

where the mean values are defined as,

$$
\bar{I}_{\text{g}} = \frac{1}{N_\phi} \sum_{\phi=1}^{N_\phi} I_{\text{g}}(\phi), \quad \bar{I}_{n,\text{scaled}} = \frac{1}{N_\phi} \sum_{\phi=1}^{N_\phi} I_{n,\text{scaled}}(\phi, t_0)
$$

{\bf From $K$ independent subaverages, each containing $n$ pulses, we obtain the following set of correlation coefficients:}

$$
\{X_n(t_0^{(1)}), X_n(t_0^{(2)}), \ldots, X_n(t_0^{(K)})\}
$$

The mean correlation and its uncertainty are given by,

$$
\langle X_n \rangle = \frac{1}{K} \sum_{i=1}^{K} X_n(t_0^{(i)}), 
$$
$$
\sigma_{X_n} = \sqrt{\frac{1}{K-1} \sum_{i=1}^{K} [X_n(t_0^{(i)}) - \langle X_n \rangle]^2}
$$

$$
\sigma_{\langle X_n \rangle} = \frac{\sigma_{X_n}}{\sqrt{K}}
$$

To study the stabilization trend, we analyze the relationship between $\log_{10}(1 - \langle X_n \rangle)$ and $\log_{10}(n)$ as shown in Figure \ref{fig:RR} for the sample MSPs.

The uncertainty in $\log_{10}(n)$ is obtained by standard error propagation:

$$ \frac{1}{\ln(10)} \cdot \frac{\sigma_{\langle X_n \rangle}}{1 - \langle X_n \rangle}
$$

We then perform a linear regression in this logarithmic space to determine the slope of the stabilization curve, which characterizes the rate at which the average profile approaches the global shape, and is presented in Table \ref{tab:1}.

{\bf For clarity of presentation, we now omit the angular-bracket notation and denote the mean $\langle X_n \rangle$ simply as $X_n$.}

\section{Results} \label{sec:results}
We present an analysis of MSP profile stability to better understand the factors influencing pulse shape variations. Cross-correlating subaveraged profiles with the global profile allows us to account for variations in intensity, width, and phase at pulse maximum.
\begin{figure*}[ht!]
\begin{center}
\includegraphics[width=1.0\textwidth,angle=0]{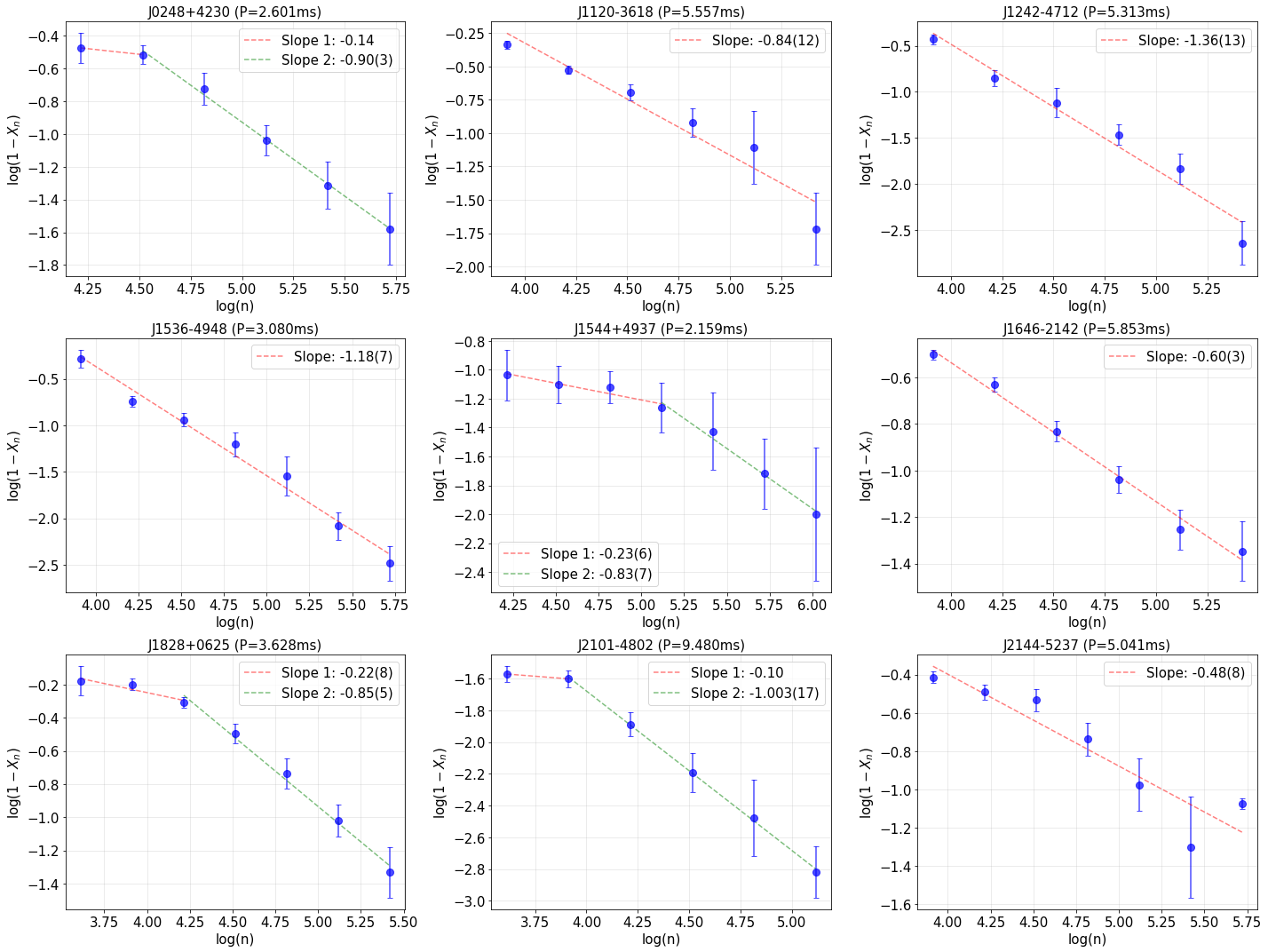}
\caption{Long-term profile stability of nine GMRT discovered MSPs using mean correlation from several years of observations. Here, $n$ denotes the number of individual pulses averaged to form a subaveraged pulse profile. For each value of $n$, we compute the Pearson correlation coefficient $X_n$ between the subaveraged profile and the global profile. The quantity plotted is $\log_{10}(1 - X_n)$ as a function of $\log_{10}(n)$, capturing the stabilization behavior of pulse profiles with increasing averaging. The vertical error bars represent the standard error of the mean of $X_n$ at each $n$, propagated through the logarithmic transformation. These correspond to 1-$\sigma$ uncertainties. The slope fitted line shows the profile stability rate.}
\label{fig:RR}
\end{center}
\end{figure*}
Figure \ref{fig:RR} displays a plot of  $\log_{10}$$(1 - X_{n})$ versus $\log_{10}$$(n)$ for all MSPs in our sample.
The stability timescale differs among MSPs. On average, achieving profile stability (a correlation of $\sim$0.99) requires averaging \(10^{5}\) to \(10^6\) periods. Consistent with the findings of HMT75 and RR95 for normal pulsars, we observe that in some MSPs (e.g. J1120$-$3618, J1242$-$4712), log$(1 - X_{n})$ varies linearly with log$(n)$, whereas some others exhibit a distinct break in the trend (e.g. J0248$-$4230, J1828$+$0625).
Following RR95, we identified three key quantities to characterize stabilization behavior in Figure \ref{fig:RR}: (1) correlation coefficient for subaverages of $n$ pulse periods, (2) the slope of the log$(1 - X_{n})$ vs. log$(n)$ line, (3) the presence of any break in the fitted line. For the GMRT discovered MSPs in our sample, it was not possible to calculate the single-pulse correlation coefficient as done by RR95 (for some of the strong normal pulsars), due to a very low single-pulse signal-to-noise ratio (SNR). We found that MSPs with broader, multi-component profiles (e.g., J1536–4948, J2144–5237, J1120–3618) exhibit slower profile stabilization, reflected in lower correlation coefficients for a given number of pulses, compared to narrow, single-component MSPs (e.g., J1242–4712), despite having similar brightness levels. A similar result was found by \citet{2019ApJ...872..193L}, who demonstrated that the jitter parameter ($k_{J}$) correlates with both the duty cycle (as a proxy for pulse width) and the number of pulse components, using the NANOGrav 12.5-year dataset. They reported a weighted correlation coefficient of $R=0.62$ between $k_{J}$ and duty cycle (W$_{50}$), and $R=0.40$ between $k_{J}$ and the number of components. A similar trend was noted qualitatively in \citet{2021MNRAS.502..407P}. Unlike normal pulsars, as observed by RR95, the observed breaks in the stability curves for MSPs could not be attributed to any specific correlated behavior like nulling and mode changing. \par

\begin{figure*}[ht!]
\begin{center}
\includegraphics[width=1.0\textwidth,angle=0]{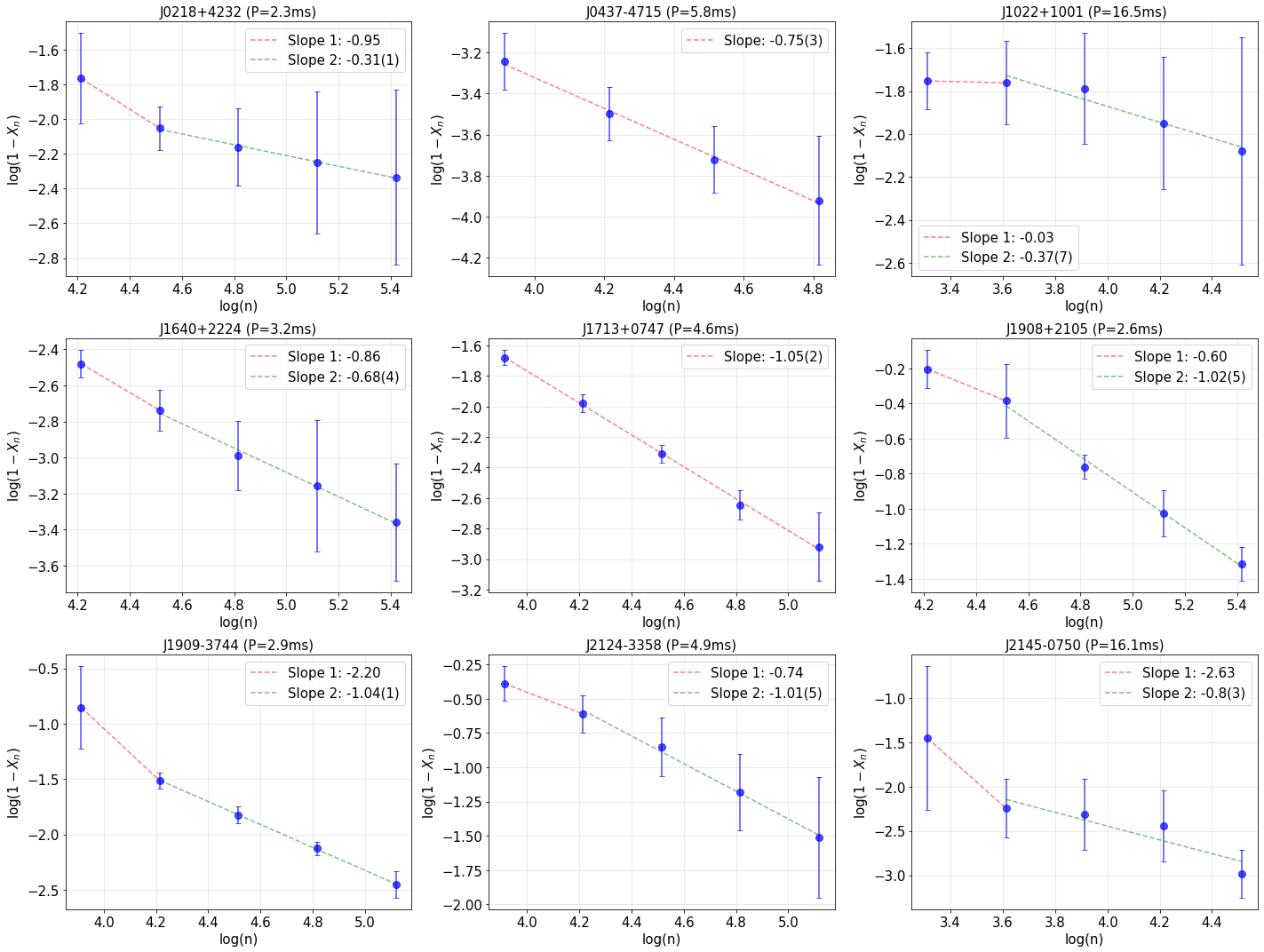}
\caption{Profile stability for a few strong MSPs observed with GMRT in a single epoch. Axis are same with Figure \ref{fig:RR}. The slope of the fitted line represents the profile stability rate.}
\label{fig:other}
\end{center}
\end{figure*}

\begin{figure*}[ht!]
\begin{center}
    \begin{minipage}{0.31\textwidth}
        \centering
        \includegraphics[width=\textwidth]{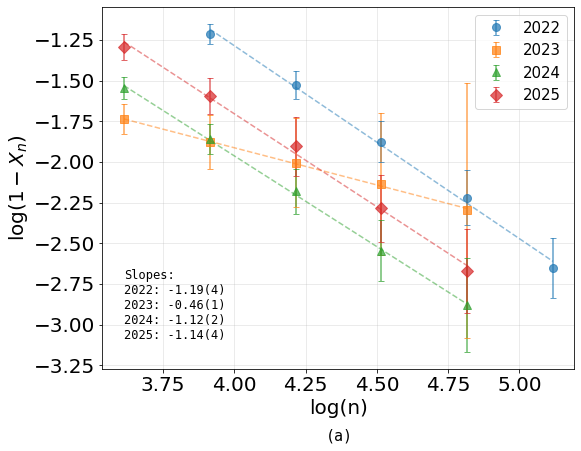}
    \end{minipage}
    \hspace{0.01 cm}
    \begin{minipage}{0.31\textwidth}
        \centering
        \includegraphics[width=\textwidth]{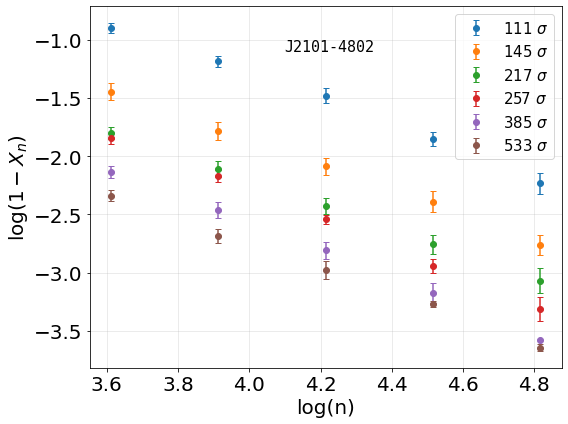}
    \end{minipage}
    \hspace{0.01 cm}
    \begin{minipage}{0.31\textwidth}
        \centering
        \includegraphics[width=\textwidth]{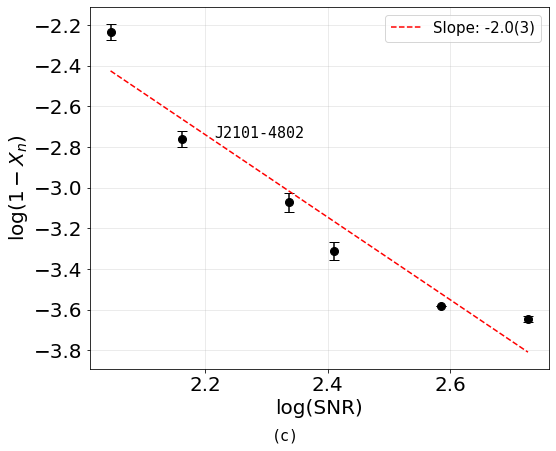}
    \end{minipage}
    \caption{(a) Profile stability of J2101$-$4802 across different years, (b) The variation in correlation with SNR, (c) The dependence of $(1 - X_n)$ on SNR are for J2101$-$4802}
    \label{fig:J2101}
\end{center}
\end{figure*}

In addition to the GMRT-discovered MSPs, we have also looked into the profile stability of nine other bright MSPs (Figure \ref{fig:other}). Out of these, J0437-4715, J1022$+$1001, J1640$+$2224, J1713$+$0747, J1909$-$3744, J2124$-$3358, and J2145$-$0750 are part of the Pulsar Timing Arrays (PTA, e.g. \citet{https://ui.adsabs.harvard.edu/link_gateway/1979ApJ...234.1100D/doi:10.1086/157593, 2016MNRAS.458.1267V, 2019MNRAS.490.4666P}). However, for these MSPs in this study, we could use only single observing epochs due to the lack of data availability, which will be addressed in the future. Though RR95 reported results from observation carried out in a single observing epoch, the observations of these bright MSPs were carried out in separate observing epochs. Among these nine MSPs, J1022$+$1001—despite being quite bright—exhibits the shallowest slope in its stability curve. 
However, as we can see, in the panel (a) and (b) of Figure \ref{fig:J2101}, the stability varies over different observing epochs, we haven't taken into account the stability of these MSPs to calculate correlation between different pulsar parameters presented in Table \ref{tab:2}.

Table \ref{tab:1} summarizes the stability analysis for all the target MSPs, listing the spin period ($P$), period derivative ($\dot{P}$), flux density at 400 MHz  ($S400$), average number of pulses per observation, the number of pulses ($n$) at which a break occurs, the slope of the stability curve before (\( S_1 \)) and after (\( S_2 \)) the break, and the classification of MSPs.\par
We also examined the profile stability across different years by calculating the mean correlation from multiple observations of similar duration throughout each year. The panel (a) of Figure \ref{fig:J2101} illustrates this year-to-year variation in the stability rate for one of our sample MSPs, J2101$-$4802. We observed differences in correlation for a given number of pulse periods across different years for all MSPs in our sample (plots for all MSPs can be found \href {https://docs.google.com/document/d/1cm-yrTF9\_u-bWBSz7z\_mH8hwXJ\_4VVUjtGTROdGdA5A/edit?usp=sharing}{\textcolor{blue}{here}}).\\
 Our analysis shows the relation $(1 - X_{n}) \propto (\text{SNR})^{-\alpha}$, where $\alpha$ varies between 1.4$-$2.2 for our sample MSPs (e.g. panel (b) and panel (c) of Figure \ref{fig:J2101} presents the same for J2101$-$4802), confirming the theoretical and simulated results presenting the relation between SNR and correlation factors by \citet{2012MNRAS.420..361L}. However, this relation is contrary to the findings of HMT75 for normal pulsars. 
 
 As noted in HMT75, the frequency dependence of the correlation value of subaverages of similar length can be attributed to two contributing factors. First, the modulation index of pulsar intensity fluctuations decreases monotonically with increasing observing frequency \citep{1975ApJ...195..513T}. Second, since the mean profile shape largely determines the strength of subpulses falling at various longitudes within the pulse window, the decreasing modulation at high frequencies implies fewer strong pulses outside profile component peaks and thus a general increase in the degree of correlation between subaverages and the mean profile. The dependence of subpulse width on observing frequency may act either to enhance or to mitigate this modulation index effect. Wider subpulses generally lead to higher correlation coefficients, while narrower subpulses tend to reduce the degree of correlation.

 \begin{figure*}[ht!]
\begin{center}
    \begin{minipage}{0.31\textwidth}
        \centering
        \includegraphics[width=\textwidth]{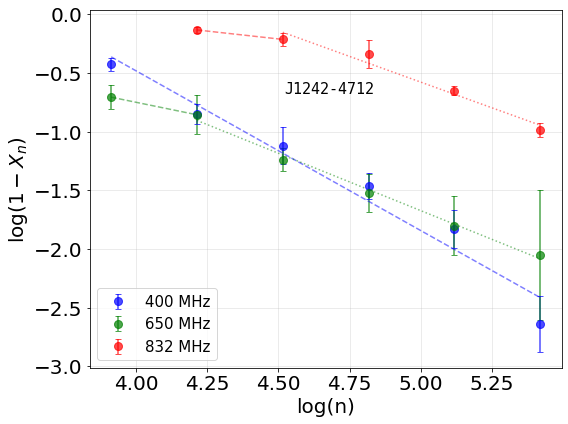}
    \end{minipage}
    \hspace{0.01 cm}
    \begin{minipage}{0.31\textwidth}
        \centering
        \includegraphics[width=\textwidth]{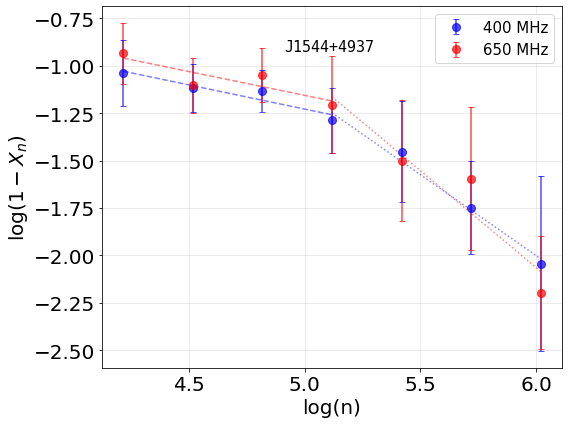}
    \end{minipage}
    \hspace{0.01 cm}
    \begin{minipage}{0.31\textwidth}
        \centering
        \includegraphics[width=\textwidth]{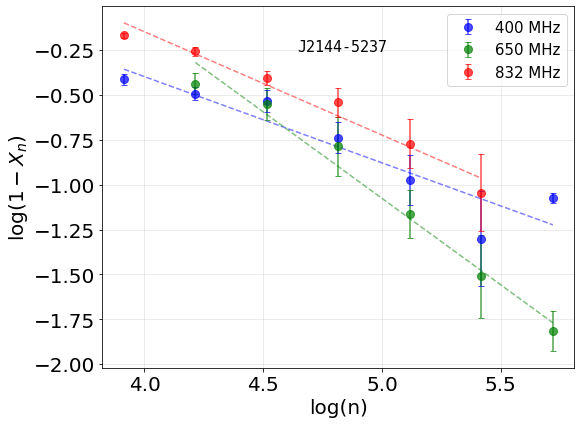}
    \end{minipage}
    \caption{Correlation plots for J1242$-$4712, J1544$+$4937, and J2144$-$4712 at different frequencies.}
    \label{fig:freq}
\end{center}
\end{figure*}
 
We analyzed frequency dependence of the profile stability rate at 400, 650, and 832 MHz (each having $\sim$200 MHz bandwidth) for MSPs J1242$-$4712, J1544$+$4937, and J2144$-$5237 (Figure \ref{fig:freq}). We found that for lower frequencies (at 400 and 650 MHz), the stability curves follow a similar trend, which is somewhat offset from the stability curve at 832 MHz, exhibiting a rather shallower slope. The additional break observed in the stability curve for J1242--4712 at 650 and 832 MHz could be attributed to the emergence of the additional profile components at these frequencies. Our study demonstrates that frequency-dependent SNR, profile evolution, and spectral characteristics all affect profile stability in MSPs, highlighting the necessity of multi-frequency observations to comprehend pulse shape fluctuations.\par

 To quantify the linear relationships between various pulsar parameters and their profile stability (quantified by stability slopes S1 and S2), we computed Pearson correlation coefficients \citep{pearson1895} along with their associated uncertainties using a Monte Carlo sampling approach. For each parameter pair, we generated 10,000 random realizations of the data by sampling from normal distributions centered on the measured values, with standard deviations equal to their respective uncertainties. The Pearson correlation coefficient was then calculated for each realization, resulting in a distribution of correlation values. The mean of this distribution is reported as the correlation coefficient, and its standard deviation represents the uncertainty (error bar) on the correlation. This method was used to account for the propagation of measurement uncertainties into the correlation analysis.
The correlations between the slopes of the log \((1-X_n)\) versus log \(n\) and the pulsar parameters for the nine GMRT discovered MSPs are summarized in Table \ref{tab:2}. The quantity \(S_{\text{Imax}}\), which represents the slope of the variation of \(I_{\text{max}}(n)-<I_{\text{max}}>\) (where \(I_{\text{max}}(n)\) is the average peak intensity of subaverages of \(n\) periods and \(<I_{\text{max}}>\)the peak intensity of the global profile), shows a strong correlation with \(\dot{P}\) and \((P\dot{P})^{1/2}\) and a strong anti-correlation with \(P/2\dot{P}\). This is similar to RR95's findings on normal pulsars, suggesting that the mechanisms causing pulse-to-pulse fluctuation in both normal pulsars and MSPs may be alike. RR95 detected a moderate anticorrelation between \(S_1\) and \(P/2\dot{P}\), indicating that older pulsars take longer to stabilize. However, we found this relationship to be weaker, suggesting that possibly profile stability is less dependent on age for MSPs.\par
RR95 discusses the possibility that oscillations of neutron stars could contribute to the irregular variations observed in pulse emission. Such as thermally induced non-radial g-mode oscillations of the neutron star crust. This oscillation timescale varies inversely with the square root of the neutron star crust's surface temperature. \citep{1989ApJ...346..808S} provide estimates of the temperature evolution as a function of P and $\dot{P}$, which leads to the oscillation timescale:  $T_g \sim \left( \frac{P^2}{\dot{P}} \right)^{1/4}$, assuming a constant moment of inertia for all neutron stars.\par
\renewcommand{\tabcolsep}{2pt} 
\begin{table*}[!htb]
\begin{center}
{\footnotesize
\caption{Correlation of stabilization with physical parameters. Values are quoted to match the precision of their uncertainties using condensed notation.}
\label{tab:2}
\begin{tabular}{lcccc} 
\toprule
Parameter & $S_{1}$ & $S_{2}$ &  $S_{I\text{max}}$ & $S_{1}-S_{2}$ \\ 
\midrule
$P$                      &  $0.04(3)$ & $0.00(5)$ &  $0.29(3)$  & $0.05(5)$ \\
$\dot{P}$                & $-0.40(5)$ & $-0.64(8)$ &  $0.77(2)$  & $-0.2(1)$ \\
$\sqrt{P \dot{P}}$       & $-0.19(5)$ & $-0.44(8)$ &  $0.74(2)$  & $0.1(1)$ \\
$\frac{P}{2\dot{P}}$     & $-0.14(8)$ &  $0.1(1)$ & $-0.40(6)$  & $-0.3(1)$ \\
$\left(\frac{P^2}{\dot{P}}\right)^{1/4}$ & $0.00(7)$ &  $0.3(1)$ & $-0.36(5)$  & $-0.2(1)$ \\
$S_{1}$                  &  --- &  $0.50(9)$ & $-0.30(6)$  & $0.77(8)$ \\
$S_{2}$                  &  $0.51(9)$ & --- & $-0.23(9)$  & $-0.1(1)$ \\
\bottomrule
\end{tabular}
}
\end{center}
\end{table*}

For normal pulsars, the negligible correlation between the stability slopes ($S$) and the oscillation timescale \(\left(\frac{P^2}{\dot{P}}\right)^{1/4}\) suggests that thermally induced oscillations do not significantly contribute to the irregular variations in pulse emission. Likewise, Alfvén modes—where the magnetic field serves as the restoring force—appear to have little influence, as indicated by the weak correlation between \(S\) and the magnetic field strength \((P\dot{P})^{1/2}\). Similarly, for MSPs, we find no significant correlation between \(S\) and the oscillation timescale \(\left(\frac{P^2}{\dot{P}}\right)^{1/4}\). However, unlike normal pulsars, MSPs exhibit a stronger anticorrelation between \(S_{I_{\text{max}}}\) and the oscillation timescale, suggesting that intensity variations in MSPs are more strongly affected by these timescales than in normal pulsars. 

Additionally, the stability slope \(S_2\) shows a moderate anticorrelation with the magnetic field strength \((P\dot{P})^{1/2}\), indicating that stronger magnetic fields may contribute to increased profile instability in MSPs. This suggests that magnetic fields might have a more significant role in the profile stability of MSPs than in normal pulsars, as observed by RR95. We also note that in giant pulse-emitting pulsars, intense magnetic fields in the outer magnetosphere are linked to the generation of bright pulses (e.g. B1821−-24 \citet{2001ApJ...557L..93R}, B1937+21 \citet{1996ApJ...457L..81C},  J1823--3021A \citet{2005ApJ...625..951K}). Moreover, the presence of such bright pulses has been shown to distort the average pulse profile and increase ToA (Time of Arrival) scatter, leading to elevated levels of jitter noise (e.g., J1713+0747, \citet{2014MNRAS.443.1463S}). Conversely, the absence of these bright pulses in most MSPs may explain their typically low jitter noise (e.g., J1909--3744, \citep{2014MNRAS.443.1463S}). This implies that MSPs with higher magnetic field may have shallower profile stability slope (longer profile stability time scale), which aligns with our finding of anticorrelation between the stability slope \(S_2\) and the magnetic field strength \((P\dot{P})^{1/2}\).

\section{Discussion on profile stability and jitter noise}

In this study, we have used multiple epochs of observations for nine GMRT-discovered MSPs for profile stability analysis. These MSPs are comparatively weak, and therefore the radiometer noise \citep{2014MNRAS.443.1463S,2016ApJ...819..155L} dominates in the timing residual, and jitter analysis becomes less effective due to the low S/N of single pulses. On short timescales, the dominant sources of white noise in pulsar timing residuals are typically radiometer noise, jitter noise, and scintillation noise \citep{2014MNRAS.443.1463S,2016ApJ...819..155L}. The total white noise in the measurement of a ToA can be expressed as:

\[
\sigma_{\text{total}}^2 = \sigma_{\text{S/N}}^2 + \sigma_{\text{J}}^2 + \sigma_{\text{DISS}}^2
\]

where \(\sigma_{\text{S/N}}\), \(\sigma_{\text{J}}\), and \(\sigma_{\text{DISS}}\) represent the contributions from radiometer noise, jitter noise, and scintillation noise, respectively.

The ratio of \(\sigma_{\text{S/N}}\) to \(\sigma_{\text{J}}\) is directly proportional to the SNR of an equivalent single pulse \citep{2012MNRAS.420..361L}. Observations with high sensitivity have shown that single pulses from nearly every pulsar exhibit variability in amplitude and phase beyond what is expected from radiometer noise alone \citep{2014MNRAS.443.1463S}. This variability introduces pulse-to-pulse shape and phase instabilities that ultimately affect the stability of the integrated profile. However, estimation of the jitter parameter can be effectively done for relatively bright sources, where single-pulse SNR is sufficiently high \citep{2012MNRAS.420..361L}. 
To address this, our study focuses on determining the profile stability timescale -- the integration time required for the pulse profile to become stable to reach a high correlation ($\sim$0.99) with the global profile, such that the influence of shape variations and jitter on ToA scatter is minimized.
For a few bright PTA MSPs with single-epoch observations, existing jitter analyses are available in the literature. We compare our profile stability results with these studies to provide a complementary perspective. \citet{2015MNRAS.449.1158L} conducted simultaneous observations of PSR J1022$+$1001 with the Westerbork Synthesis Radio Telescope (WSRT) and the Effelsberg 100-m Radio Telescope and observed that pulse shape variations showed excess scatter in ToAs, larger than expected from only radiometer noise. They found that the integrated profile varies on a time scale of a few tens of minutes. \citet{2021MNRAS.502..407P} reported that the modulation index, which measures intensity variations from pulse to pulse, grows increasingly stronger across the pulse profile. All of these findings support our observation of a slower stability rate and reduced correlation between subaverages for J1022$+$1001.
In contrast, although \citet{2011MNRAS.418.1258O, 2014MNRAS.443.1463S} and \citet{2021MNRAS.502..407P} reported significant pulse shape variations in MSP J0437$-$4715, this MSP shows the highest degree of correlation between subaverages and the mean profile for a given number of pulses. Being the brightest in our sample, this is consistent with the expectation that profile stability improves with increasing SNR, as discussed in Section \ref{sec:results}.
\citet{2021MNRAS.502..407P} reported wide variability in phase and amplitude in the single pulses of MSP J2145$-$0750, leading to significant levels of jitter noise. They also reported the first observation of pulse nulling in MSP J1909−-3744, which is notably one of the MSPs exhibiting the lowest levels of jitter noise. However, in our single-epoch observation, the phase and amplitude variations in J2145--0750 appear less prominent, though they show a comparatively slower stabilization rate. Moreover, we do not find evidence for nulling in MSP J1909$-$3744.
\renewcommand{\tabcolsep}{4pt}
\begin{table*}[!htb]
\begin{center}
{\footnotesize
\caption{Slope of stabilization ($S_2$) and jitter noise estimates from Lam et al. (2019) for five bright MSPs. Jitter values are shown for Model A and Model C. The corresponding jitter parameter $k_J = \sigma_J / P$ is also listed.}
\label{tab:jitter}
\begin{tabular}{lccccccc}
\toprule
Pulsar & $P$ (ms) & $S_2$ & $\sigma_{J,A}$ ($\mu$s) & $\sigma_{J,C}$ ($\mu$s) & $k_{J,A}$ & $k_{J,C}$ \\
\midrule
J1022$+$1001 & 1.645 & --0.37(7) & 381.9$\pm$5.2 & 343.0$\pm$7 & 0.2322(32) & 0.2085(43) \\
J1640$+$2224 & 3.163 & --0.68(4) & 37.0$\pm$1.9 & 11.3$\pm$1.8 & 0.0117(6) & 0.0035(5) \\
J1713$+$0747 & 4.570 & --1.05(2) & 51.0$\pm$0.2 & 69.4$^{+0.6}_{-0.5}$ & 0.0111(4) & 0.0152(1) \\
J1909$-$3744 & 2.947 & --1.04(1) & 16.1$\pm$0.4 & 20.5$^{+0.5}_{-0.6}$ & 0.0054(1) & 0.0069(1) \\
J2145$-$0750 & 1.605 & --0.8(3) & 338.4$^{+3.0}_{-3.1}$ & 329.3$^{+3.9}_{-5.2}$ & 0.2108(19) & 0.2050(28) \\
\bottomrule
\end{tabular}
}
\end{center}
\end{table*}

To quantitatively assess the connection between the stability metrics and jitter noise, we compared the slope $S_2$ from our profile stabilization analysis with the jitter parameter $k_J = \sigma_J / P$ reported in \citet{2019ApJ...872..193L} for five PTA MSPs. As shown in Table \ref{tab:jitter}, we find a strong correlation between the stabilization slope and the value of $k_J$, with Pearson correlation coefficients of --0.72 (p-value 0.16) and --0.67  (p-value 0.21) for Models A and C, reported in  \citet{2019ApJ...872..193L}. This suggests that pulsars with higher levels of intrinsic jitter noise tend to show a slower rate of profile stabilization. We note that the slope parameters ($S_1$, $S_2$, reported in Table \ref{tab:1}) are derived from averaged profiles and cannot be directly extrapolated to single-pulse behavior. In this work, we demonstrate a high correlation between profile stability slope and jitter parameter, providing direct evidence that for weaker MSPs, where jitter analysis is not possible, profile stability analysis can give an indirect way of assessing the intrinsic pulse shape variability.

This approach is particularly beneficial for MSPs for which the direct jitter measurement is not feasible because of low single pulse SNR. With this approach, we could additionally explore the correlation of profile stability rate with the intrinsic pulsar parameters such as characteristic age, spin period, period derivative, and magnetic field strength, providing insights into whether any of these properties play a significant role in driving profile shape instability.

\section{Summary and Future scope} \label{sec:summary and conclusions}
This study presents the first systematic investigation of pulse profile stability in MSPs and its dependence on pulsar properties, demonstrated for long-term observations of nine MSPs discovered with the GMRT. We find that the timescale required to achieve profile stability varies with observing frequency and is strongly influenced by SNR and pulse morphology. 

Single-epoch high-SNR observations of some bright MSPs using uGMRT band 3 reinforce that stabilization timescales are strongly influenced by profile shape and intrinsic brightness. Our results imply that, on average, achieving profile stability (a correlation of $\sim$0.99) requires averaging \(10^{5}\) to \(10^6\) periods, depending on pulsar brightness, observing frequency, and magnetospheric characteristics. For MSPs with complex, multiple-component profiles or higher magnetic field strengths, longer integrations may be necessary to form a stable profile. These results demonstrate that the spectral properties of MSPs--through their influence on frequency-dependent SNR, and intrinsic profile evolution significantly affect the rate at which pulse profiles stabilize. Consequently, longer observation lengths are typically required at higher frequencies to achieve similar levels of profile stability.

These findings provide critical constraints on the minimum averaging time required to mitigate jitter noise, an essential factor in precision timing, and thus have direct implications for improving timing residuals in PTA experiments. Our study uncovers a strong link between the profile‑stability slope and the jitter parameter, showing that for faint MSPs where direct jitter measurements are impractical, profile‑stability analysis can reliably stand in as an indicator of intrinsic pulse‑shape variability. This analysis opens a new window for characterizing jitter in MSPs, particularly those too faint for conventional jitter measurements.

Constraints on the minimum timescale for achieving a stable pulse profile in MSPs with different pulse shapes can help optimize jitter noise and thus improve timing residuals, thereby enhancing the sensitivity of PTA. We note that MSPs show a stronger anticorrelation between intensity variations and neutron star crust oscillation timescales compared to normal pulsars. The observed anticorrelation between intensity fluctuations and crustal oscillation timescales, along with a moderate link between profile stability and magnetic field strength, suggests a deeper connection between magnetospheric dynamics and emission stability in MSPs.  Extending this analysis to larger samples will be vital for decoding MSP emission physics and optimizing timing strategies for gravitational wave detection.

\section{Acknowledgements}
We acknowledge the support of the Department of Atomic Energy, Government of India, under project no.12-R\&D-TIFR-5.02-0700. 
The data presented in this paper were obtained by the uGMRT and the Parkes radio telescope. The Parkes radio telescope is part of the Australia Telescope, which is funded by the Commonwealth of Australia for operation as a National Facility managed by CSIRO. We also thank our referee, Michael Lam, for insightful comments that improved the quality of the paper. We thank Maura McLaughlin for helpful discussions and feedback on the analysis, and Simon Johnston for valuable guidance on the use of Parkes UWL data and for his continued support throughout this project.

\bibliography{RR.bib}{}
\bibliographystyle{aasjournalv7}

\end{document}